\documentclass{aa}
\usepackage{graphicx}
\usepackage{txfonts}
\newcommand{\Mpch}{\,{\rm Mpc}\,\ifmmode h^{-1}\else $h^{-1}$\fi}
\newcommand{\Msh}{\,\ifmmode M_\odot\,h^{-1}\else $M_\odot\,h^{-1}$\fi}

\newcommand{\hmpc}{\,$h^{-1}$\,Mpc}
\newcommand{\symmA}{Symmetron A}
\newcommand{\symmB}{Symmetron B}
\newcommand{\symmC}{Symmetron C}
\newcommand{\symmD}{Symmetron D}

\begin{document} 

\title{Degeneracies between modified gravity and baryonic physics}


   \author{Thor A. S. Ellewsen
          \and
          Bridget Falck
          \and
          David F. Mota
          }

        \institute{Institute of Theoretical Astrophysics, University of                 Oslo, Postboks 1029, 0315 Oslo, Norway\\
        \email{t.a.s.ellewsen@astro.uio.no}\\
    \email{bridget.falck@astro.uio.no}\\
    \email{d.f.mota@astro.uio.no}
        }
   \date{Received 12 September 2017 / Accepted 3 March 2018}

\abstract{In order to determine the observable signatures of modified gravity theories, it is important to consider the effect of baryonic physics. We used a modified version of the ISIS code to run cosmological hydrodynamic simulations in order to study degeneracies between modified gravity and radiative hydrodynamic processes. One of the simulations was the standard $\Lambda$CDM model and four were variations of the Symmetron model. For each model we ran three variations of baryonic processes: nonradiative hydrodynamics; cooling and star formation; and cooling, star formation, and supernova feedback. We constructed stacked gas density, temperature, and dark matter density profiles of the halos in the simulations, and studied the differences between them. We find that both radiative variations of the models show degeneracies between their processes and at least two of the three parameters defining the Symmetron model.
}

   \keywords{cosmology: dark energy -- methods: numerical -- hydrodynamics}

\maketitle
%

\section{Introduction}
Ever since the discovery of the accelerated expansion of the universe~\citep{riess1998observational,perlmutter1999measurements}, there has been a continuous effort to explain why this is happening. One explanation is that it is the product of an as-yet-unknown phenomenon dubbed dark energy. The alternative is that the theory of gravity is not entirely correct, and needs to be modified. This has resulted in numerous theories of modified gravity (MG)~\citep{carroll2005cosmology,nojiri2006modified,nojiri2007introduction,nojiri2011unified,skordis,clifton2011modified}. There are restrictions on the modification that can be made to general relativity (GR); because of the extensive tests that have been done on GR~\citep{terukina2014testing,euclid,wilcox2015xmm} and because  it has passed with flying colors, any MG theory must reduce to GR on solar system scales. This is usually achieved by a  screening mechanism~\citep{kimura2012vainshtein}. The form of the mechanism depends on the modification to the theory. For scalar-tensor theories,  multiple screening mechanisms have been proposed. Examples are Chameleon~\citep{khoury2004chameleon}, Dilaton~\citep{gasperini2001quintessence}, Vainshtein~\citep{vainshtein1972problem}, Disformal \citep{tomi}, and Symmetron~\citep{hinterbichler2010screening}. In this paper we focus on the Symmetron model.

To determine the effects of modified gravity on the  nonlinear formation of structure, $N$-body simulations are required. In addition to solving the standard force equations, modified gravity codes must also solve for the nonlinear evolution of an extra scalar field, involving unique numerical challenges. 
There have been efforts to simulate various models of modified gravity~\citep{hans,oyaizu2008nonlinear1,li2009structure,schmidt2009self,li2011n,li2012ecosmog,brax2013systematic,claudio,llinares2013releasing,falck2014vainshtein}. The main codes are summarized and tested against one another in~\citet{winther2015modified}, which found broad agreement within a few percentage points.

However, there is a major pitfall in such an approach. In order to test these theories against observations we need to compute real observables, which  cannot be directly calculated from dark matter-only simulations, since experiments only measure photons, which are emitted from the baryonic matter. 
This raises several questions. What observables from the simulations would be best suited for comparison to observations in order to put stronger constraints on modified gravity theories and to test Einstein's general relativity? Are there degeneracies between modified gravity and baryonic physics at cluster and galaxy scales? 

The  past few decades have seen a dramatic improvement in our ability to include hydrodynamic processes in $N$-body simulations, thanks to a rapid increase in computing power coupled with the development of sophisticated numerical techniques. 
At this point there are numerous simulation codes that include baryons in some form~\citep{schaye2014eagle,enzo2005introducing,gadget2005cosmological,teyssier2002cosmological}. Most codes focus on these effects within the $\Lambda$CDM framework; however, developers of modified gravity codes are beginning to include hydrodynamics as well~\citep{puchwein2013modified,hammami2015cosmological,hammami2015hydrodynamic,arnold2014scaling}. In those works, cluster properties such as halo profiles and probability distribution functions have been computed, and lately the gas-fraction of the galaxy clusters and power-spectra have been suggested as viable candidates.

Building on the work of~\citet{hammami2017probing}, who studied the mass-temperature relation of galaxy clusters, we look at the effect of including radiative processes such as cooling, star formation, and supernova feedback in simulations of the Symmetron model.
This allows us to investigate the possibility that these baryonic effects may be degenerate with the effects of modified gravity. If this is the case, it could mean that drawing conclusions about the correct gravity theory based only on dark matter properties could be misleading. Moreover, the misunderstanding of the potential degeneracies between modified gravity and baryonic effects could lead to false detections. In this paper, we study these degeneracies by looking at the density profiles of halos, for both dark matter and baryons, and at their temperature profiles.
\section{Methods}
\subsection{Theory}\label{sec:Symmetron}
The Symmetron model, first introduced by~\citet{hinterbichler2010screening}, is a special case of a scalar-tensor theory of gravity with action
 \begin{align}
 S &= \int d^4x\sqrt{-g}\left[\frac{R}{2}M_{pl}^2 - \frac{1}{2}\partial^i\psi\partial_i\psi - V(\psi)\right] \label{Sym_action} \\
 &\quad+ S_m(\tilde{g}_{\mu\nu},\tilde{\Psi}_i), \nonumber
 \end{align}
where $g$ is the determinant of the metric tensor $\tilde{g}_{\mu\nu}$, $R$ is the Ricci scalar, $\psi$ is the scalar field, and $V(\psi)$ the potential.
The potential is 
\begin{align}
V(\psi) = V_0-\frac{1}{2}\mu^2\psi^2 + \frac{1}{4}\lambda\psi^4,
\end{align}
where $\mu$ is a mass scale and $\lambda$ a dimensionless parameter. 
We consider a model with coupling
\begin{equation}
A(\psi) = 1 + \frac{1}{2M^2}\psi^2 + \mathcal{O}\bigg(\frac{\psi^4}{M^4}\bigg).
\end{equation}
To make physical interpretations easier we define
\begin{align}\label{sym_params}
 \beta  &= \frac{M_{pl}\psi_0}{M^2}, \\
a_{\rm{SSB}}^3 &= \frac{3H_0^2\Omega_m M_{pl}^2}{M^2\mu^2}, \\
\lambda_0^2 &= \frac{1}{2\mu^2},
\end{align}
as done by~\citet{winther2012environment}. In this context $\beta$ is the strength of the fifth force, $a_{\rm{SSB}}$ the scale factor at symmetry breaking, and $\lambda_0$ the maximum range of the scalar fifth force in units of \hmpc.
The fifth force is~\citep{hammami2015hydrodynamic}
\begin{equation}
F_\psi = 6\Omega_m H_0^2\frac{(\beta\lambda_0)^2}{a_{\rm{SSB}}^3}\chi\nabla\chi,
\end{equation}
where
\begin{equation}\label{fifth}
\tilde{\psi} = a\psi = \psi_0\chi.
\end{equation}

\subsection{Simulation}
We used a modified version~\citep{hammami2015hydrodynamic} of the \texttt{ISIS}~\citep{llinares2014isis} simulation code, which in turn is based on the adaptive-mesh code \texttt{RAMSES}~\citep{teyssier2002cosmological}, allowing for hydrodynamic simulations of modified gravity. 

The simulations were run using the cosmological parameters defined in Table~\ref{tab:cosmo_params}, with a box size of 64 \hmpc\ on each axis and $256^3$ dark matter particles, using six levels of refinement, with $l_{min}=8$ and $l_{max}=14$. Details on the refinement criterion and strategy used can be found in  the ISIS~\citep{llinares2014isis} and RAMSES~\citep{teyssier2002cosmological} papers.
All initial conditions were  identical for all the simulations. The box had periodic boundary conditions. The initial conditions were generated as Gaussian random fields using \texttt{GRAFIC2}~\citep{bertschinger2001multiscale,prunet2008initial} starting at  redshift $z = 49$. All the results in this paper comes from analyzing the $z=0$ snapshot of the simulations.

\begin{table}
        \centering
    \caption{Cosmological parameters used for all simulations}
    \label{tab:cosmo_params}
        \begin{tabular}{l l}
    \hline
    Parameter & Value \\
    \hline
    $\Omega_m$ & 0.3\\
    $\Omega_b$ & 0.05 \\
    $\Omega_k$ & 0 \\ 
    $\Omega_\Lambda$ & 0.7 \\
    h & 0.67\\
    \hline
        \end{tabular}
\end{table}

We used the Amiga Halo Finder~\citep{gill2004evolution,knollmann2009ahf} to find the locations of the halos and the corresponding radius, $r_{200}$, and mass, $M_{200}$. Here both $M_{200}$ and $r_{200}$ are with respect to the critical density of the universe. We kept all halos with 100 or more particles. This resulted in somewhere between 4,000 and 5,000 halos depending on the simulation and its parameters.

We ran three variations of hydrodynamic effects for both $\Lambda$CDM and Symmetron (see next section). The simulation parameters are given in Table~\ref{tab:hydro_params}.
Star formation is modeled in the same way as in the \texttt{RAMSES} code~\citep{dubois2008onset}, and the parameters were chosen to match those used in their paper. This uses the standard approach used by many works, where  a source term is added to the continuity equation, such that
\begin{equation}
\dot{\rho}_\star =  -\frac{\rho}{t_\star} ~\mathrm{if}~ \rho > \rho_0~,~ \dot{\rho}_\star =  0 ~ \mathrm{otherwise,}
\end{equation}
where $\rho$ is the gas density, $\rho_\star$ is the density of stars, and $\rho_0$ is the star formation density threshold.
The star formation timescale  is defined as
\begin{equation}
t_\star = t_0\bigg(\frac{\rho}{\rho_0}\bigg)^{-1/2},
\end{equation}
where $t_0$ is a parameter usually in the range 1--10 Gyr. If a gas cell is eligible for star formation, a Poisson random process is applied to create $N$ collisionless star particles, according to
\begin{equation}
P(N) = \frac{\lambda_p}{N!}\exp(-\lambda_p),
\end{equation}
with mean value
\begin{equation}
\lambda_p = \bigg(\frac{\rho\Delta x^3}{m_\star}\bigg)\frac{\Delta t}{t_\star}.
\end{equation}

The supernovae feedback model consists of thermal heating of the ISM as well as modeling of a blast wave. The procedure used is described in Appendix A of \citet{dubois2008onset}.

\begin{table}
        \centering
\caption{Combinations of parameters defining the hydrodynamic effects. Here $t_0$ is a timescale in Gyr, $\rho_0$ is the star formation density threshold in H $\mathrm{cm^{-3}}$, cooling turns on or off the cooling/heating term in the energy equation, and $\eta_{SN}$ controls the fraction of stars that explode into supernovae.}
    \label{tab:hydro_params}
    \begin{tabular}{l l l l l}
    Simulation  & cooling &$t_0$ & $\rho_0$ & $\eta_{SN}$\\
    \hline
    vanilla      &      False    & 0  &0  &0   \\
    cool star    &      True     & 8.0&0.1&0   \\
    cool star sn &      True     & 8.0&0.1&0.1 \\
    \hline
        \end{tabular}
\end{table}

\subsection{Baryonic effects}
One of the complications of including baryons in simulations is  that we have to model hydrodynamic and radiative effects. This complicates the physics greatly. In this work we   include cooling, star formation, and supernova feedback in the simulations; additionally, we run simulations with nonradiative hydrodynamics only, dubbed \textit{vanilla}. There are of course more phenomena that affect the physics, such as metallicity, super massive black holes (SMBH), and feedback from active galactic nuclei (AGN), but we have chosen not to include them here. 

The main effect of introducing cooling is that the temperature of the gas cools down, decreasing the thermal pressure, which in turn lets the gravitational collapse of the gas proceed more effectively. This leads to halos with higher densities in their centers than when cooling is not included. An interesting feature of cooling is that the cooling rate increases as the temperature decreases, leading to a larger fraction of gas having cooled at $z=0$ than observations show. This leads to what is known as the cooling catastrophe~\citep{balogh2001revisiting}. It is then clear that cooling alone is not enough to explain the behavior of the baryons.
Increased star formation is a consequence of increased cooling~\citep{gerritsen1997star}. However, more stars give more radiative heating, slowing down the collapse of the gas, and subsequent star formation.

The logical next step is to include the fact that a fraction of stars go supernova. This releases an enormous amount of energy back into the surrounding gas, heating it, and thus preventing collapse. We also note  that supernovae have an effect on the metallicity of the gas, which affects cooling (see~\citealt{somerville2015physical} for a nice review of these processes). 
It is clear that all the baryonic effects have an impact on each other in some form, making it complicated when deciding which ones to include. 
Recent works point towards SMBHs and AGNs playing an important role as well~\citep{martizzi2012effects}.
It would be interesting to include these in the simulations with modified gravity. Unfortunately, a proper treatment of AGN feedback would require using a larger box and higher resolution, which at this point is too computationally expensive.
For details on the implementation of the baryonic effects in the simulation code we refer the reader to the original RAMSES paper~\citep{teyssier2002cosmological}. 

In the following, \textit{vanilla} refers to the nonradiative hydrodynamic simulation; \textit{cool star} includes cooling and star formation; and \textit{cool star sn} includes cooling, star formation, and supernova feedback.

\section{Results: $\Lambda$CDM versus Symmetron}\label{sec:LcdmSymAonly}
An effective way to compare structure in simulations is to make radial profiles of various quantities for the halos in the simulations. This can be any physical quantity of interest. In this paper we look at the density of gas and dark matter, as well as their temperature profiles. The Symmetron model has a series of free parameters that need to be chosen in order to compare them to $\Lambda$CDM. For this section we use $\beta = 1$, $a_{\mathrm{SSB}} = 0.5$, and $\lambda_0  = 1$.

\subsection{Halo mass functions}\label{sec:symahmf}
\begin{figure}
        \includegraphics[width=\columnwidth]{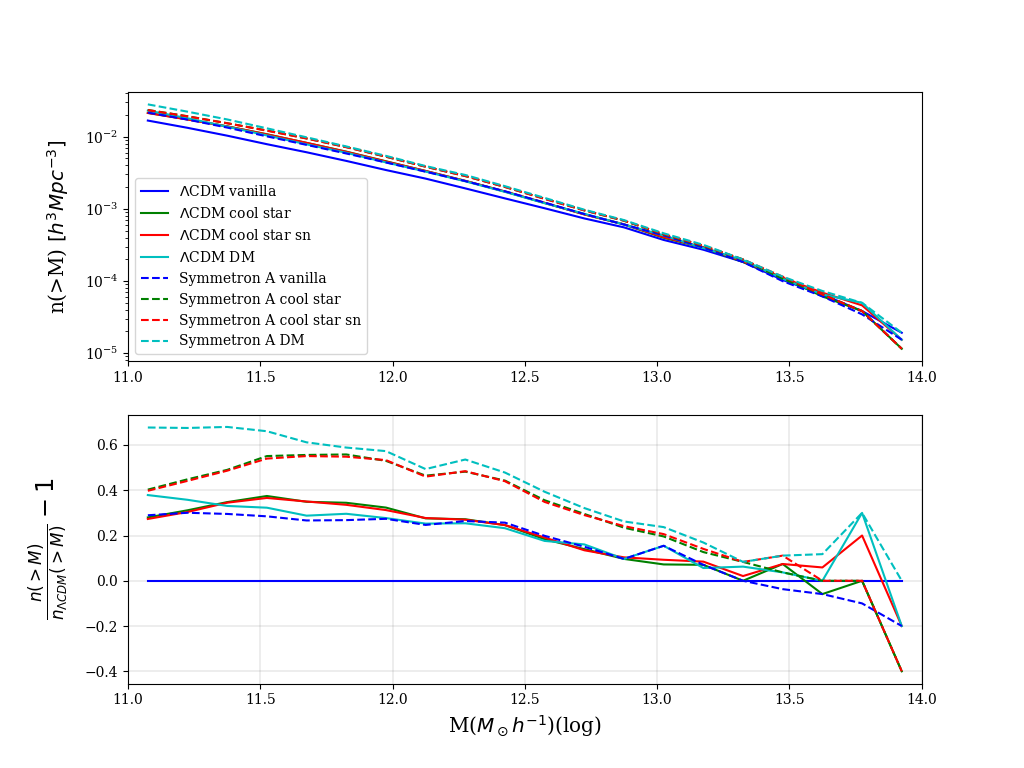}
    \caption{Top: Halo mass functions of the different simulations. Bottom: Ratio of the HMF of the simulation to the $\Lambda$CDM \textit{vanilla} HMF .}
    \label{fig:HMF_symaonly_20}
\end{figure}

We use the $M_{200}$ masses of the halos provided by the Amiga Halo Finder to make halo mass functions (HMFs). From the HMFs (Fig.~\ref{fig:HMF_symaonly_20}) we see that all the Symmetron simulations produce more low-mass  halos than their $\Lambda$CDM counterparts. This is in agreement with the findings of \citet{davis2012structure}. We also see that including supernova feedback has little effect as the HMFs of \textit{cool star} and \textit{cool star sn} overlap for both $\Lambda$CDM and Symmetron. This can be seen more clearly in the lower panel of Fig.~\ref{fig:HMF_symaonly_20}. In this figure we also note the large variations in the ratios for the higher mass ranges. This is caused by the smaller number of halos in this range, making the ratio change more with the addition of just a few halos in either simulation.

For the \textit{vanilla} versions of the simulations, the increased number of halos in the Symmetron simulations can be easily explained by the increased gravity from the fifth force, which effectively pulls in more matter and creates more halos. It is not immediately obvious that this should be the case for the \textit{cool star} and \textit{cool star sn} variations. 

\subsection{Gas density profiles}
To study degeneracies we make halo gas density profiles for each of the simulations. The profiles are calculated radially and scaled to their $r_{200}$ for all the halos within the mass range of each simulation, and then averaged. The profiles are measured from $0.1r_{200}$ to $10r_{200}$. It is worth considering whether a radius of $0.1r_{200}$ is converged or not. \citet{power2003inner} looked into this question and the many factors that need to be taken into account. While they did conclude that their results were reliable for radii larger than a few percent of the virial radius, we should still be careful when looking at small halos.

We focus on two mass ranges: $1)$ {\it{large halos}} with masses larger than $10^{13.5}\Msh$;  these halos are usually screened and the fifth force is small; $2)$ {\it{small halos}} with masses smaller than $10^{12}\Msh$;  these halos are usually not screened and the fifth force is large. This gives us about 4,000 halos for the low-mass group, and about $30$ for the high-mass group. For the mass range between the two chosen, we expect there to be a mix of screened and unscreened halos, making it harder to draw clear conclusions from them. There are, however, between 500 and 1,000 halos in this range which may be useful for studies at a later point.

Since we want to look for degeneracies, we plot the ratios between the gas density profiles of $\Lambda$CDM and Symmetron for each of the variations of hydrodynamics. 
For the halos with masses $10^{13.5}\Msh$ and higher  (Fig.~\ref{fig:lcdmsymaonlyfigs}) we see that the screening is working at scales smaller than $r_{200}$, making the deviations from $\Lambda$CDM small. 

For the halos with masses $10^{12}\Msh$ and lower (Fig.~\ref{fig:lcdmsymaonlyfigs}), we see that including cooling and star formation does not have the same effect in the Symmetron model as in $\Lambda$CDM. If there was no degeneracy, all three lines would overlap. Since this is not the case we can conclude that cooling and star formation is degenerate with the Symmetron model. Furthermore, we  see that introducing supernova feedback does not change the ratio between Symmetron and $\Lambda$CDM. Thus, we can also conclude that supernova feedback is not degenerate with Symmetron. We note that this relies on the choice of parameters for the Symmetron model not being special in any way. We  look at this in section~\ref{sec:symmcomp}.

There is an interesting point to be made about the difference between Symmetron and $\Lambda$CDM when cooling and star formation is introduced. As can be seen from Fig.~\ref{fig:lcdmsymaonlyfigs}, the gas density of the Symmetron halos is significantly lower in the centers of halos than that of  $\Lambda$CDM. This goes against the assumption that cooling would increase gravitational collapse. The cause of this is not clear at the moment. It is possible that the halo population in the low-mass range is skewed toward the lower end in Symmetron making the average profile of this mass range lower. Nevertheless, this does not change the conclusion that there is a degeneracy.

\subsection{Temperature profiles}
To calculate the temperature of the halos we use the ideal gas law
\begin{equation}
p = R_s \rho T,
\end{equation}
where $p$ is the thermal pressure, $R_s=k_b/\bar{m}$ is the specific gas constant, $\bar{m}= 0.59m_H$ is the mean mass of the gas, $m_H$ is the hydrogen mass, $\rho$ is the gas density, $k_b$ is the boltzmann constant, and $T$ is the temperature. Since the profiles are very similar, we  look instead at the ratio of the Symmetron model to its corresponding $\Lambda$CDM variation. We also note  that the profiles are mass weighted such that $T = \sum m_i T_i /\sum m_i$.

\begin{figure*}
        \includegraphics[width=\textwidth]{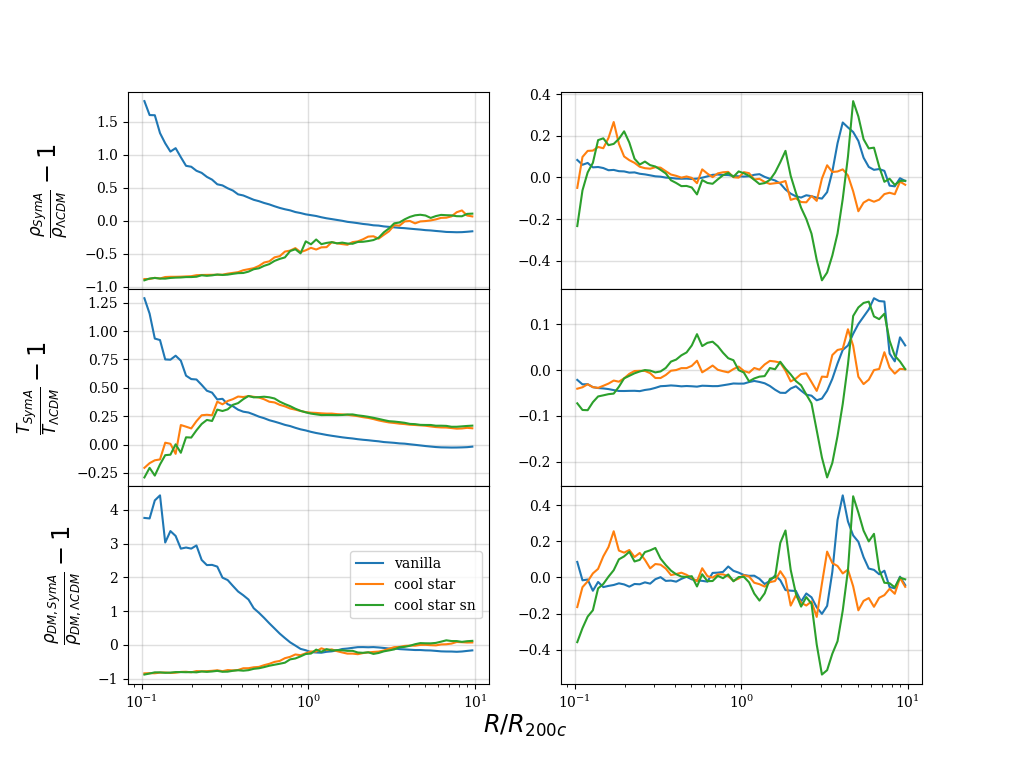}
    \caption{Left: Halos with mass lower  than $10^{12}\Msh$. Right: Halos with mass higher than $10^{13.5}\Msh$. Top: Gas density. Middle: Gas temperature. Bottom: Dark matter density.}
    \label{fig:lcdmsymaonlyfigs}
\end{figure*}

We  started by looking at the screened group ($M>10^{13.5}\Msh$) (Fig.~\ref{fig:lcdmsymaonlyfigs}). 
Recall that \textit{vanilla} is the variation that includes baryons but no radiative processes, \textit{cool star} includes cooling and star formation, and \textit{cool star sn} includes supernova feedback as well. From the figure it becomes clear that the variations between the models are small, as is expected; since the screening mechanism is effective, all of them behave close to general relativity. Including radiative effects does little to change this because the halos are very large;  therefore, the impact of the baryonic effects is less relevant.

For the halos with masses smaller than $10^{12}\Msh$ (Fig.~\ref{fig:lcdmsymaonlyfigs}) we see a different behavior.
First, the \textit{vanilla} curve shows that Symmetron has a higher temperature at the centers of halos than $\Lambda$CDM. Second, after introducing cooling and star formation, this is no longer the case. Instead Symmetron now has a lower temperature near the center. Finally, introducing supernova feedback has the same effect on both $\Lambda$CDM and Symmetron. 
This supports the results of the gas density profiles.

\subsection{Dark matter density profiles}\label{sec:SymAonlyDM}
Considering that we are working with baryonic effects, we wanted to see whether the dark matter is significantly affected by this through their gravitational interaction with the baryons. To study this we plotted radial dark matter density profiles for the simulations. As before, we looked at the average profile for each group of halos in the same mass ranges. 

We started with the higher than $10^{13.5}\Msh$ group (Fig.~\ref{fig:lcdmsymaonlyfigs}). Just as we saw for gas density and temperature, the dark matter density profile ratios also show small deviations between the models. While the introduction of cooling, star formation, and supernova feedback does change the profile, there is no clear evidence that they have an effect.

For the lower than $10^{12}\Msh$ group (Fig.~\ref{fig:lcdmsymaonlyfigs}) we get another picture. Looking at the \textit{vanilla} versions we see that Symmetron has a higher dark matter density than $\Lambda$CDM at radii smaller than $r_{200}$. With the introduction of cooling and star formation the situation flips, and $\Lambda$CDM has the higher density in the centers. In addition, as we saw with the gas density, the supernova feedback gives the same change to both Symmetron and $\Lambda$CDM. 

This is the same behavior that we saw for the gas density. Since cooling and star formation does not affect dark matter directly we are left with two possible reasons for this. Either the baryons have successfully pulled the dark matter with it, or we are simply looking at a significantly different halo population. Either way, this also supports the conclusion that there is a degeneracy between cooling and star formation and Symmetron.

\section{Results: Symmetron comparison}\label{sec:symmcomp}
We saw in Section~\ref{sec:LcdmSymAonly} that switching from $\Lambda$CDM to Symmetron had an impact on the effect of hydrodynamics on halo profiles. However, it is not clear which parameter of the Symmetron this is connected to. To make this clear, we have to look at different variations of Symmetron.
As explained in Section~\ref{sec:Symmetron}, there are three parameters that control the model:  the coupling strength $\beta$, the scale factor at which symmetry breaking happens $a_{\mathrm{SSB}}$, and the maximum range of the fifth force $\lambda_0$. We  chose to use three variations of Symmetron in addition to the one already studied, denoted \symmA. 

From eq.~\ref{fifth} we see that the three parameters of the Symmetron model work together to define the strength of the fifth force. Here we see that both increasing $\beta$ and reducing $a_{\mathrm{SSB}}$ results in a stronger fifth force, meaning that it may be possible to achieve the same effect using different sets of parameters. We note that $\lambda_0$ remains the same for all variations since we expect there to be a degeneracy between the three parameters, making it pointless to vary them all.

The parameters of the models can be seen in Table~\ref{tab:model_params}. 
\begin{table}
        \centering
        \caption{Four Symmetron models and their corresponding parameters.}
        \label{tab:model_params}
        \begin{tabular}{l l l l}
        \hline
        Symmetron model & $\beta$ & $a_{\mathrm{SSB}}$ & $\lambda_0$\\
        \hline 
        Symmetron A & 1.0 & 0.5  & 1.0\\
        Symmetron B & 1.0 & 0.33 & 1.0\\
        Symmetron C & 2.0 & 0.5  & 1.0\\
        Symmetron D & 1.0 & 0.25 & 1.0\\
        \hline
        \end{tabular}
\end{table}
To see the differences between the models, we used the same methods as in Section~\ref{sec:LcdmSymAonly}, but this time we compared the different Symmetron models. 

\subsection{Halo mass functions}
\begin{figure}
        \includegraphics[width=\columnwidth]{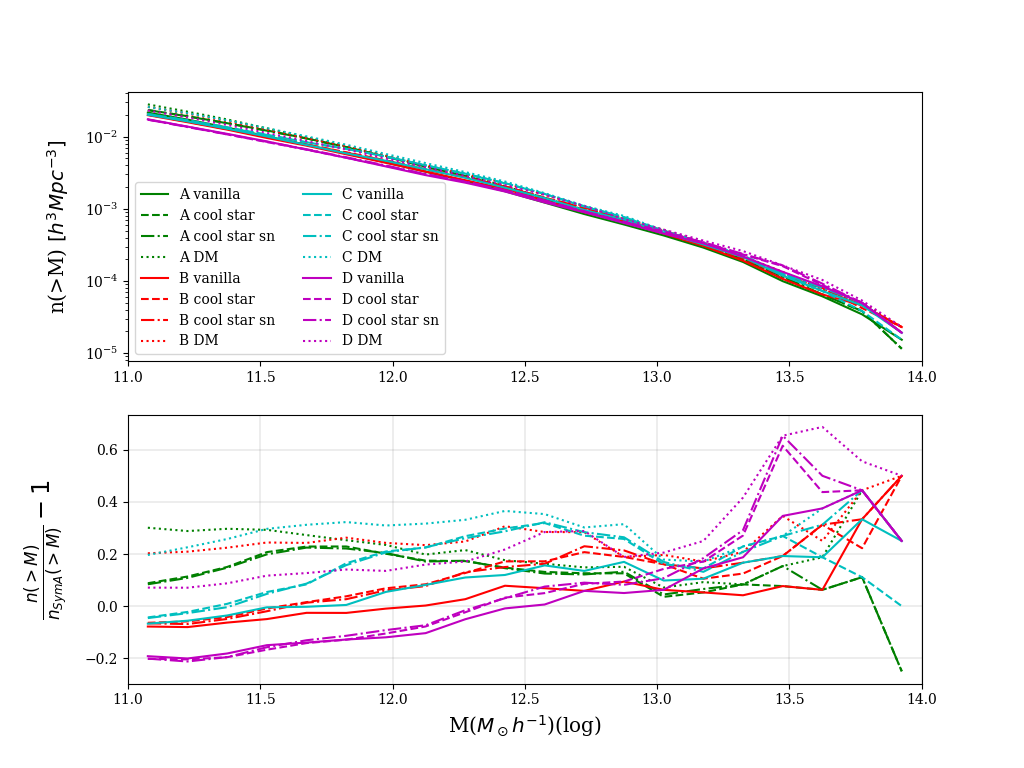}
    \caption{Top: HMFs of the different simulations. Bottom: Ratio of the HMF of the simulation to the \symmA\ \textit{vanilla} HMF.}
    \label{fig:HMF_nolcdm_20}
\end{figure}

We start by looking at the halo mass functions as before. As we saw in section~\ref{sec:symahmf}, there is a lot of overlap, which makes  it  more interesting to study the ratios to \symmA\ \textit{vanilla} (Fig.~\ref{fig:HMF_nolcdm_20}). From this we see that there is less variation in the HMFs  between the Symmetron models than we saw between \symmA\ and $\Lambda$CDM. We also note that reducing $a_{\mathrm{SSB}}$ decreases the number of low-mass halos. In fact, reducing $a_{\mathrm{SSB}}$ reduces the total number of halos overall. \citet{davis2012structure} studied this for dark matter only $N$-body simulations. They looked at the mass range $10^{12}\Msh$ to $10^{14}\Msh$ and found that both stronger coupling strength, $\beta$, and earlier symmetry breaking resulted in more low-mass halos. We find the same for $\beta$, but for the $10^{12}\Msh$ to $10^{13}\Msh$ mass range we do not observe the same behavior with respect to $a_{\mathrm{SSB}}$. Whether this is because small halos are absorbed into larger halos, because of an effect of the hydrodynamics, or something else is unclear. 
Considering the hydrodynamic effects, we see that introducing cooling and star formation causes an increase in halos larger than $10^{12}\Msh$ for all models and that including supernova feedback has almost no effect on this. We also note that the variations in the HMF ratios for high masses are caused by the relatively small number of halos in this range making the ratios sensitive to small changes. 

\subsection{Gas density profiles}
As before we start with high-mass halos where the screening mechanism should be active (Fig.~\ref{fig:symcompfigs}). Recall that \symmA\ has $a_{\mathrm{SSB}} = 0.5$, while \symmB\ has $a_{\mathrm{SSB}} = 0.33$, and \symmD\ has $a_{\mathrm{SSB}}=0.25$. Also recall that \symmC\ has $\beta = 2$ instead of $1$ like the rest. 

The first thing we see is that all the simulations give similar profiles for radii smaller than $r_{200}$. The \textit{vanilla} curves for \symmB\ and \symmD\ tell us that reducing $a_{\mathrm{SSB}}$ results in less matter in the centers of halos. An interesting feature of $a_{\mathrm{SSB}}$ is what happens at radii larger than $r_{200}$. Going from \symmA\ to \symmB\ results in an increase in density with a peak around $2\,r_{200}$, while going further to \symmD\ results in a decrease. We note, however, that both models have lower densities than \symmA\ at higher radii, with their low points at approximately $4\,r_{200}$. \symmC\ does not show any deviations from \symmA\ at radii smaller than $r_{200}$; however, it does  result in a decrease in density around $3\,r_{200}$, and an increase at $5\,r_{200}$.

Introducing cooling and star formation (dashed lines), results in deviations from the \textit{vanilla} lines for all three models, indicating a degeneracy between cooling/star formation and both $a_{\mathrm{SSB}}$ and $\beta$. For \symmB\ this results in a deviation from \symmA\ of almost $1.75$ times the density at approximately $2\,r_{200}$. \symmD\ does the same with an increase of almost $2.75$ times the density of \symmA\ at the same radius. The change in \symmC\ is not as dramatic. Instead of deviating further from its \symmA\ counterpart, it instead results in a  profile that is similar to the  \symmA\ profile.

Adding supernova feedback to the models has approximately the same effect on  \symmB\ and \symmD\ as it does on \symmA, meaning we see no indication of a degeneracy between $a_{\mathrm{SSB}}$ and supernova feedback. For \symmC\ there is an increase in density around $3r_{200}$ compared to the corresponding \symmA\ model, indicating a degeneracy between $\beta$ and supernova feedback.

For the group of halos with masses lower than $10^{12}\Msh$ (Fig.~\ref{fig:symcompfigs}) the picture is quite different. 
For the \textit{vanilla} models we see that \symmC\ has a higher density than \symmA\ at all scales, while \symmB\ and \symmD\ have a lower density than \symmA\ for $r<r_{200}$ and a higher density than \symmA\ for $r>r_{200}$. We also note that the difference between the \symmB\ and \symmD\ curves shows that decreasing $a_{\mathrm{SSB}}$ results in an increase in density at all radii. We note, however, that both of these models have $a_{\mathrm{SSB}}$ smaller than \symmA\ with which we take the ratio, meaning that even though going from $a_{\mathrm{SSB}}= 0.33$ to $0.25$ gives an increase in density at the centers of halos, we also know that going from $0.5$ to $0.33$ caused a decrease.

The \textit{cool star} variations (dashed lines) show a decrease in density on scales smaller than $r_{200}$ for \symmB, and a larger decrease for \symmD. The introduction of cooling and star formation makes the difference between the two models change the other way. For \symmC\ there is a decrease in density around $0.7r_{200}$.

With the introduction of supernova feedback we see that the variations  from cooling and star formation are even smaller. The most interesting change is in the \symmC\ models with supernova feedback. Here we see that the combination of supernova feedback and a larger $\beta$ (the coupling strength of the Symmetron field to matter) results in a density  approximately 9 times higher than that of the \symmA\ equivalent at the centers of these large halos. 

\subsection{Temperature profiles}
As in the previous sections, we start by studying what happens to the $10^{13.5}\Msh$ group (Fig.~\ref{fig:symcompfigs}). 

The first thing to note is the increase in temperature for all models. The only deviation from this is \symmD, whose temperature is  approximately $1.2$ times higher  at $r < r_{200}$ compared to that of \symmA, and  $0.9$ times lower at around $4\,r_{200}$.

Turning on cooling and star formation in these models results in an increase in temperature for \symmC\ of approximately $1.2$ times that of  \symmA\ at $r> r_{200}$. This supports the conclusion from the gas densities that there is a degeneracy between $\beta$ and cooling/star formation. We reach the same conclusion with respect to $a_{\mathrm{SSB}}$. This is  clearest from the \symmD\ profiles. Instead of maintaining the $1.2$ times increase compared to \symmA,\, we now see an increase of as much as $1.6$ times.

As before, we do not see any change in the profiles of \symmB\ and \symmD\ when including supernova feedback in the simulations. There is a slight increase in temperature for \symmC\ between $3\,r_{200}$ and $4\,r_{200}$, supporting the conclusion of a degeneracy between $\beta$ and supernova feedback.

Next we look at the $<10^{12}\Msh$ group (Fig.~\ref{fig:symcompfigs}).
For the \textit{vanilla} curves we see that \symmB\ generally has a higher temperature than \symmA. \symmD\ has an even higher temperature, and here the increase goes all the way to the center of the halo. The same happens for \symmC, but here we get the same temperature increase at all radii, instead of a larger increase at larger radii than at small ones, as is seen for  \symmB\ and \symmD . 

The \textit{cool star} curves show an increase in temperature for all models. We see that \symmB\ has the smallest increase relative to \symmA. \symmD\ behaves in  the same way as \symmB, but with a higher temperature overall. \symmC\ also has an increase at all scales, but the increase in temperature is higher at the center of the halo relative to its \symmA\ counterpart than for the \textit{vanilla} version. Thus, we can again conclude that cooling/star formation is degenerate with both $a_{\mathrm{SSB}}$ and $\beta$.

The \textit{cool star sn} variants all exhibit a larger increase in density from the introduction of supernova feedback than the \symmA\ version. As before, \symmB\ shows the least increase of the three models. \symmD\ behaves the same as \symmB, just scaled up. \symmC\ and \symmD\ with supernova feedback reach a temperature almost 9 times higher than \symmA\ does at the center of the halos. Thus we can also conclude that supernova feedback is degenerate with both $a_{\mathrm{SSB}}$ and $\beta$.

At radii larger than $r_{200}$, we see the curves separate into bands according to their Symmetron model. It is interesting to see that the difference between the models comes down to their Symmetron parameters at large radii where the hydro effects become largely irrelevant. 

\subsection{Dark matter density profiles}
\begin{figure*}
        \includegraphics[width=\textwidth]{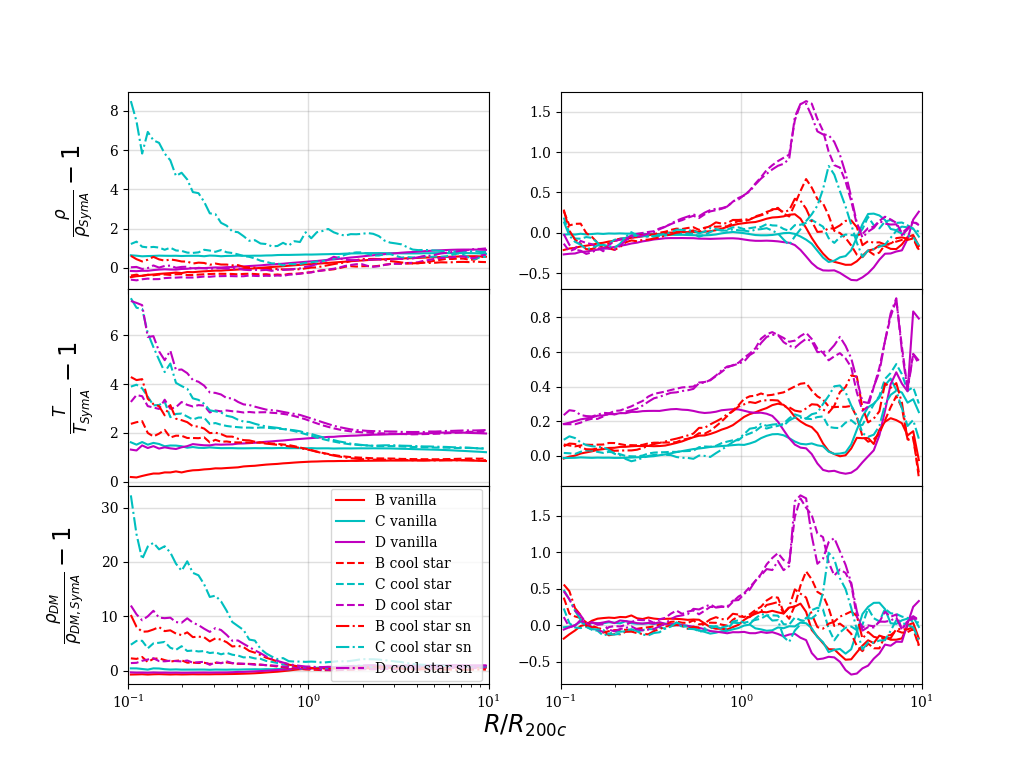}
    \caption{Left: Halos with mass lower  than $10^{12}\Msh$. Right: Halos with mass higher  than $10^{13.5}\Msh$. Top: Gas density. Middle: Gas temperature. Bottom: Dark matter density.}
    \label{fig:symcompfigs}
\end{figure*}

The dark matter density profiles for the higher than $10^{13.5}\Msh$ group (Fig.~\ref{fig:symcompfigs}) show practically identical profiles to the gas density. Thus, the discussion of the profiles leads to the same conclusions, supporting what we have seen from both the gas density and the temperature profiles.

We next study the low-mass group ($M< 10^{12}\Msh$) in Fig.~\ref{fig:symcompfigs}.
The \textit{vanilla} models all lie close to the \symmA\ model. 
Switching to the \textit{cool star} variants we see that \symmB\ and \symmD\ have switched places, indicating that the effect of cooling and star formation on the dark matter profiles are affected by the change in $a_{\mathrm{SSB}}$. We also note that \symmC\ has an increased density as well, indicating that $\beta$ also affects cooling and star formation. We note that these are baryon only effects, and we are looking at dark matter profiles, meaning that these effects are the result of the baryons pulling on the dark matter gravitationally. 
Finally, the \textit{cool star sn} models show that changing $a_{\mathrm{SSB}}$ and $\beta$ also affects supernova feedback. Changing $\beta$ from $1$ to $2$ results in an approximately 30 times increase in density for \symmC\ at the centers of halos compared to \symmA, indicating a strong degeneracy between supernova feedback and the strength of the fifth force $\beta$.

\section{Conclusions}
We have investigated the possibility of degeneracies between baryonic processes and modified gravity by running $N$-body simulations with three variations of baryonic processes in $\Lambda$CDM and the Symmetron model. From studying the differences between $\Lambda$CDM and the Symmetron model we found that there is a degeneracy between modified gravity and hydrodynamics. 

We found this by studying stacked density profiles for two mass scales: one with masses higher than $10^{13.5}\Msh$, where we expect the screening mechanism to be active, and one with masses lower than $10^{12}\Msh$, where we expect the screening mechanism to be inactive.

When studying the low-mass halos, we saw that introducing cooling and star formation had a significant effect on the gas density profiles, and that its effect was different for Symmetron than for $\Lambda$CDM.
For the high-mass halos, we did not see a significant difference in the models when introducing radiative effects. Studying stacked profiles for gas temperature and dark matter density supported this.

To better understand degeneracies between the hydrodynamic effects and the modified gravity model, we ran simulations with different model parameters and compared these simulations to the standard Symmetron model. The parameters we chose to vary were the scale factor at which the symmetry break happens, $a_{\mathrm{SSB}}$, and the strength of the fifth force, $\beta$.

We first looked at the screened group, and found that there were degeneracies between $a_{\mathrm{SSB}}$ and cooling and/or star formation, but not between $a_{\mathrm{SSB}}$ and supernova feedback. Changing $\beta$ did not affect the behavior of cooling and/or star formation significantly, and neither did supernova feedback; however, we did  see small variations, indicating a possible degeneracy. The temperature and dark matter density profiles agreed with this as well.

From studying the low-mass halos we saw that the change from introducing cooling and star formation was affected by the variations in $a_{\mathrm{SSB}}$. Furthermore, the effect of introducing supernova feedback was also affected by the variation of the parameter, indicating a degeneracy between $a_{\mathrm{SSB}}$ and both cooling and/or star formation and supernova feedback for the unscreened group. By varying $\beta$ we saw that this also affected the change caused by introducing cooling and star formation. With supernova feedback included there was an even greater difference between the models. This indicates a degeneracy between $\beta$ and both cooling and/or star formation, and supernova feedback, where the degeneracy is stronger for the latter.
The dark matter density profiles agreed with this.
From the temperature profiles of this group we drew the same conclusions. We note, however, that the temperature profiles indicate that the degeneracy between the parameters and the baryonic effects happen at scales smaller than some approximate radii, with the \textit{cool star} and \textit{cool star sn} models separating from the \textit{vanilla} models at $r \approx 2r_{200}$, and the \textit{cool star sn} model separating from the \textit{cool star} model at at $r \approx 0.6r_{200}$. 

In short, when looking at the high-mass halos we saw a clear degeneracy between $a_{\mathrm{SSB}}$ and cooling/star formation, and indications of a degeneracy between $\beta$ and supernovae feedback. When considering the low-mass halos we saw degeneracies between both radiative variations and $a_{\mathrm{SSB}}$ as well as $\beta$. 

We also studied the halo mass functions of the simulations, and found that the Symmetron model predicts more halos, giving different halo populations for the models. We posit that this could be the explanation for the counterintuitive results we saw for the density profiles. If this is the case, we may have to tune the parameters of the baryonic effects differently depending on the parameters of the modified gravity. This also pointed toward there being a degeneracy between the modified gravity and the baryonic effects.

The logical next step would be to include treatment of metallicity, SMBHs, and AGNs, as well as varying the parameters controlling the various baryonic effects. We plan to perform these investigations in a future work.

\begin{acknowledgements}
We thank the anonymous referee for the many insightful comments, Amir Hammami for letting us use his code, and the yt-project community for assistance with visualizing the data. We also thank the Research Council of Norway for their support. The simulations used in this paper were performed on the NOTUR cluster {\tt{HEXAGON}}, which is the computing facility at the University of Bergen. This paper is based upon work from COST action CA15117 (CANTATA), supported by COST (European Cooperation in Science and Technology). 
\end{acknowledgements}


\bibliographystyle{aa}
\bibliography{biblio}

\end{document}